\documentclass[twocolumn,preprintnumbers,
               superscriptaddress]{revtex4}
\usepackage{epsfig, fancybox}

   \newcommand{\be}{\begin{equation}}
   \newcommand{\ee}{\end{equation}}

   \newcommand{\bea}{\begin{eqnarray}}
   \newcommand{\eea}{\end{eqnarray}}

  \newcommand{\bm}[1]{\mbox{\boldmath$#1$}}

  \newcommand{\veps}{\varepsilon}

\begin{document}

\title{Quark Spectrum above but near Critical
Temperature  of Chiral Transition}
\author{Masakiyo Kitazawa}
\affiliation{Yukawa Institute for Theoretical Physics,
Kyoto University, Kyoto 606-8502, Japan}
\affiliation{Institute fuer Theoretische
Physik, J.W.Goethe-Universitaet,D-60054 Frankfurt am Main, Germany}
\author{Teiji Kunihiro}
\affiliation{Yukawa Institute for Theoretical Physics,
Kyoto University, Kyoto 606-8502, Japan}
\author{Yukio Nemoto}
\affiliation{Department of Physics, 
Nagoya University, Nagoya, 464-8602 Japan}

\begin{abstract}
We explore the quark properties at finite temperature 
near but above 
the critical temperature  of the chiral phase transition. 
We investigate the effects of the precursory soft mode of the
phase transition on the quark dispersion relation and the 
spectral function.
It is found that there appear novel excitation spectra of quasi-quarks
and quasi-antiquarks with a three-peak structure, 
which are not attributed to
the hard-thermal-loop approximation.
We show that the new spectra originate from the 
mixing between a quark (anti-quark) and an anti-quark hole
(quark hole) caused by a ``resonant scattering'' 
of the quasi-fermions with the thermally-excited soft mode which has
a small but finite excitation energy.
\end{abstract}

\date{\today}
\maketitle


\section{Introduction}  \label{intro}
It has been  being revealed that 
the QCD matter above
the chiral and deconfined phase transition 
at high temperature ($T$) 
seems unexpectedly  rich in physics.
The data from Relativistic Heavy-Ion Collider (RHIC) at
BNL\cite{Arsene:2004fa}, for instance, 
indicate that the matter produced in this region shows robust
collective flows, which can not be explained by
the perturbative QCD\cite{Shuryak:2003xe}.
Analyses of the elliptic flow at low $p_T$
suggest the created QCD matter has 
an almost vanishingly small viscosity\cite{Teaney:2003pb},
which can be naturally understood only when  
the created matter near the critical temperature ($T_c$)
is a strongly coupled system.
Recent lattice simulations of QCD, though in the quenched level,
 also suggested that the lowest charmonium states 
seem to survive well above $T_c$ as spectral peaks
\cite{Asakawa:2003re}; 
namely, the hadronic charmonium states may have
a longer life than might  be originally expected\cite{Matsui:1986dk}.

In short, one may say that {\em the condensed matter physics of the
QCD matter in the high-temperature phase} of the QCD phase transitions
is now making a form.
We should like notice here that such a condensed matter physics
was already anticipated some twenty years ago\cite{Hatsuda:1985eb};
the existence of hadronic elementary excitations above $T_c$
was suggested for the first time for the light quark sector,
on the basis of the chiral symmetry and the assumption
that the phase transition is a second order or weak first 
order. According to the wisdom obtained in the
condensed matter physics,
if the phase transition is such a type, then
one can expect the existence of specific elementary excitations
which are coupled to the fluctuations of the order parameter.
These excitation modes are known in condensed matter physics as
soft modes associated with the phase transition.
Since the soft modes above $T_c$ in the chiral transition have 
the same quantum number
as the sigma meson and the pion, so they may called the 
`para-sigma($\sigma$) meson' 
and the `para-pion($\pi$)'\cite{orsay}; see also 
\cite{Schafer:1995df}
for  the instanton liquid approach.
In this Letter, we show that the quark spectra just above 
$T_c$ of the chiral transition also are quite different from the 
free quark spectra and show an interesting behavior, as a
critical phenomenon of the phase transition.

It has been shown very recently 
\cite{Kitazawa:2003cs,Kitazawa:2004cs2,Kitazawa:2005pp}
that the precursory {\em diquark} fluctuations\cite{Kitazawa:2001ft} are developed
 so greatly to form a soft mode at high density but at moderate temperature
that the quark spectrum is also significantly modified by dressing the
soft mode;
there arises a depression of the quark spectrum around the
Fermi surface leading to the {\em pseudogap} in the density of states (DOS)
of quarks.

It is thus highly expected that the precursory
soft modes of the chiral phase transition
should also strongly affect the quark spectrum near $T_c$.
In this Letter, we investigate how such soft modes composed of the light 
quarks affect in turn the quark properties, i.e, the dispersion relations
and the spectral function\cite{Kitazawa:2005cg}.
Needless to say,  it is important to know the properties of 
quarks (and gluons) for studying, for example, the possible formation of
hadronic bound states of quarks,
as is done in \cite{Shuryak:2004cy,Brown:2003km,Mannarelli:2005pz}, too.
We shall show that the  coupling between a
quark and a hole of the thermally excited anti-quarks\cite{Weldon:1989ys} 
becomes significant 
through the coupling with the thermally excited soft modes, which 
leads to an interesting complications to the quark spectrum. 
We also give an intuitive account for the formation of such a spectra
and the relation to the plasmino 
spectrum found at extremely high temperatures\cite{Klimov,Weldon:1989ys,Bellac}.

Here we consider the chiral limit ($m_u=m_d=0$) in the two flavors
which leads to the second order phase transition at low density.
This is because we can study the fluctuation effects to the quark
spectrum genuinely in this case and the finite quark mass effects
may make the mechanism of the quark spectrum more complicated.
The analysis with finite current quark masses is left as a future work.
We shall also confine ourselves to the case of the vanishing chemical
potential in this paper.

\section{Soft modes and quark spectral function} \label{form}

To describe the quark matter near $T_c$,
we employ the two-flavor Nambu--Jona-Lasinio (NJL) model
\cite{Nambu:1961tp}
\be
  \mathcal{L}=\bar{\psi} i \partial \hspace{-0.5em} / \psi
  + G_S [(\bar{\psi} \psi)^2 + (\bar{\psi}i\gamma_5\vec{\tau}\psi)^2],
\ee
as an effective model of low-energy QCD\cite{Hatsuda:1994pi}
with $\vec{\tau}$ being the flavor SU(2) Pauli matrices.
The coupling constant $G_S=5.5$ GeV${}^{-2}$ and the three dimensional
cutoff $\Lambda=631$ MeV are taken from Ref.~\cite{Hatsuda:1994pi}.
This model gives the second order transition at $T_c=193.5$MeV.
In the following discussions, we limit our attention to
the system at $T$ higher than  $T_c$
 where chiral symmetry is restored.

Since the basic ingredient for the following discussions is the 
existence of the soft modes associated with the chiral transition
at  $T$ above but near $T_c$ \cite{Hatsuda:1985eb},
we first recapitulate the results in \cite{Hatsuda:1985eb} with
some elaboration needed for the subsequent  discussions.

We denote the quark-antiquark retarded Green functions in the 
scalar and pseudo-scalar channels
as $D_\sigma^R(\bm{p},\omega)$ and $D_\pi^R(\bm{p},\omega)$
 with the subscripts $\sigma$ and $\pi$,
because  they have the same quantum numbers as the $\sigma$ meson and
the pion, although they are excitations in the Wigner phase of the
chiral symmetry. The collective modes in these channels are called
`para-$\sigma$' and `para-$\pi$', respectively. 
The $D_\sigma^R(\bm{p},\omega)$ and $D_\pi^R(\bm{p},\omega)$ are
obtained in the imaginary-time formalism:
The corresponding  imaginary-time propagators
in the random phase approximation(RPA) 
read
\be
  \mathcal{D}_\sigma(\bm{p},\nu_n)= \mathcal{D}_\pi(\bm{p},\nu_n)=
  -\frac{2G_S}{1+2G_S \mathcal{Q}(\bm{p},\nu_n)},
\ee
where 
$\nu_n=2\pi n T$ is the Matsubara frequency for bosons and 
$\mathcal{Q}(\bm{p},\nu_n)$ is the one-loop quark-antiquark 
polarization function,
\be
  \mathcal{Q}(\bm{p},\nu_n)=T\sum_m \int \frac{d^3 q}{(2\pi)^3}
  {\rm Tr} [ \mathcal{G}_0(\bm{q},\omega_m)
  \mathcal{G}_0(\bm{p}+\bm{q},\nu_n+\omega_m) ],
\ee
where
$
  \mathcal{G}_0(\bm{p},\omega_n)=[ i\omega_n\gamma_0
  -\bm{p}\cdot \bm{\gamma}]^{-1}
$ is the free quark propagator 
with $\omega_n=(2n+1)\pi T$ being the Matsubara frequency for fermions;
the trace is taken over color, flavor and Dirac indices.

With the standard  analytic
continuation, we have  the retarded Green functions,
$D_{\sigma,\pi}^R(\bm{p},\omega)=
\mathcal{D}_{\sigma,\pi}(\bm{p},\nu_n)|_{i\nu_n=\omega+i\eta}$.
The spectral functions in the scalar and the pseudo-scalar channels are given by
\be
 \rho_{\sigma,\pi}(\bm{p},\omega) 
= -\frac{1}{\pi}{\rm Im}D^R_{\sigma,\pi}(\bm{p},\omega),
\ee
respectively.

\begin{figure*}
\begin{center}
\begin{tabular}{ccc}
$\varepsilon = 0.1$ & $\varepsilon=0.2$ & $\varepsilon=0.5$ \\
\epsfig{file=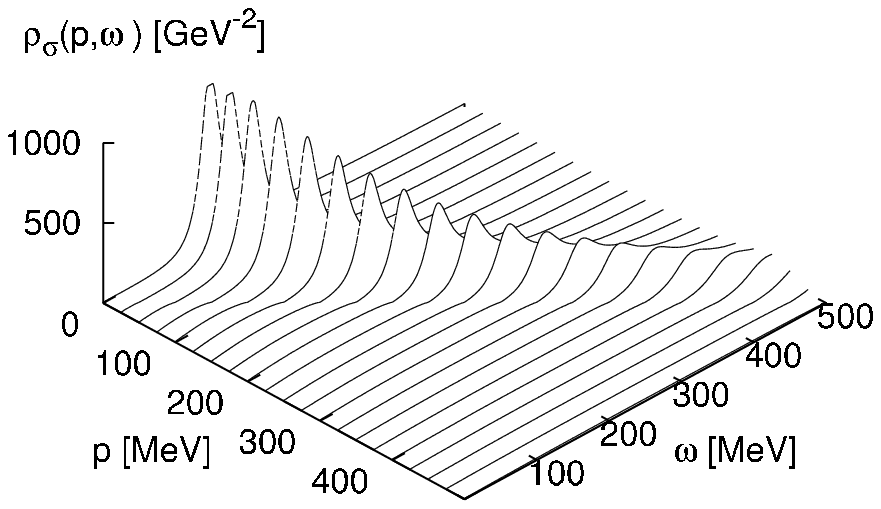, width=.32\textwidth}&
\epsfig{file=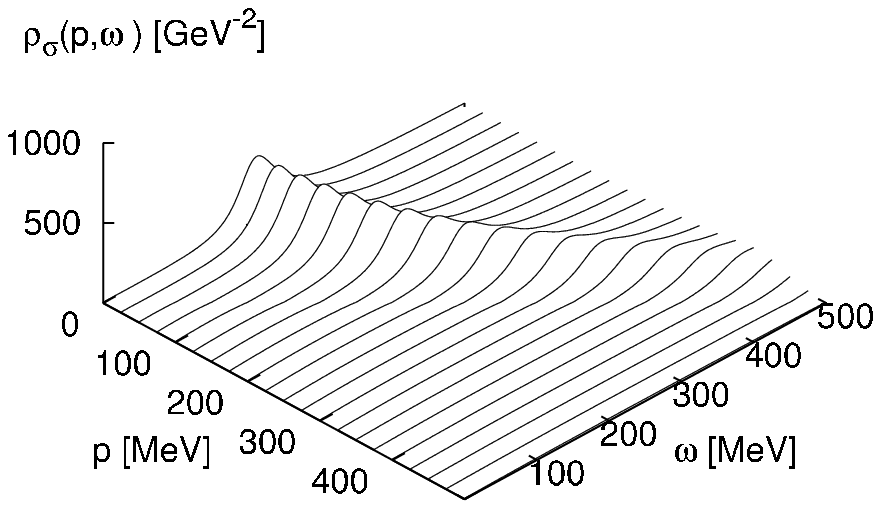, width=.32\textwidth} &
\epsfig{file=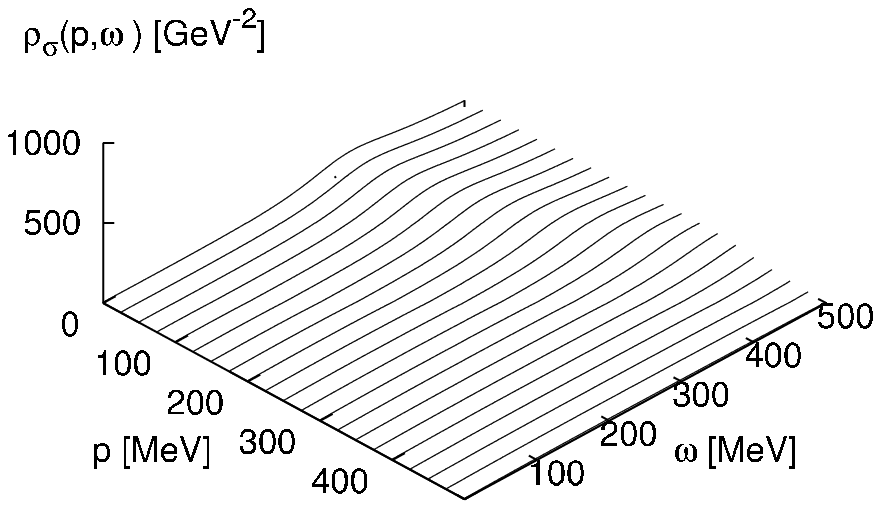, width=.32\textwidth}
\end{tabular}
\caption{
The spectral function of the `para-$\sigma (\pi)$' mode $\rho_{\sigma,\pi}$
as a function of the energy and momentum for the reduced temperature
$\varepsilon =0.1,\, 0.2$ and 0.5, from the far left to the right.
When $T$ is close to $T_c$, there appear
well developed peaks yielding the existence of collective modes or 
 well defined elementary excitations. As $T$ goes high away from $T_c$
the peaks become obscure and eventually disappear.
}
\label{fig:softmode}
\end{center}
\end{figure*}

For later convenience,
 we show in Fig.~\ref{fig:softmode} the spectral function 
$\rho_{\sigma,\pi}(\bm{p},\omega)$ of the `para-$\sigma (\pi)$' mode
as a function of the energy $\omega$  and 
momentum $\bm{p}$ for some  reduced temperatures 
$\varepsilon\equiv (T-T_c)/T_c$.
Although not shown here,
the spectral function $\rho_{\sigma,\pi}(\bm{p}, \omega)$ has
a strength for negative energies also and satisfies the symmetry property;
 $\rho_{\sigma,\pi}(\bm{p}, -\omega)=-\rho_{\sigma,\pi}(\bm{p}, \omega)$: 
One can see that 
there appears a pronounced peak in $\rho_{\sigma,\pi}(\bm{p},\omega)$
in the time-like region.
The peak position can be expressed approximately as
$\omega \simeq \pm\sqrt{ m^*_\sigma(T)^2 + |\bm{p}|^2 }
\equiv \pm\omega_s(\bm{p};T)$
with a $T$-dependent `mass'  $m^*_\sigma(T)$,
and as $T$ approaches $T_c$ the width of the peaks and $m^*_\sigma(T)$
become smaller, which means that there exist the soft modes for 
the chiral transition, as was first shown in \cite{Hatsuda:1985eb}.

Some remarks are in order here:
(1)~The present soft modes
are a propagating mode with a finite frequency or `mass'
$m^*_{\sigma}(T)$ even at $\bm{p}=0$, in contrast to
those in the (color-)superconductivity where 
the soft mode is almost  diffusive with
a large strength around $\omega=0$\cite{Kitazawa:2004cs2}.
This difference will be found to cause a quite different behavior 
for the dressed quark spectra.
(2)~The fact that the spectral function $\rho_{\sigma,\pi}$ has 
sharp peaks around 
$\omega=\pm\omega_s(\bm{p};T)$ at $T$ close to $T_c$
means that the soft modes may be well described 
as elementary scalar and pseudo-scalar fields with 
the mass $m^*_\sigma(T)$ in this $T$ region.
(3)~The `para-$\sigma$($\pi$)' mode has a strength
mostly in the time-like region,
although there is a tiny strength
also in the space-like region.

\begin{figure}
\begin{center}
\epsfig{file=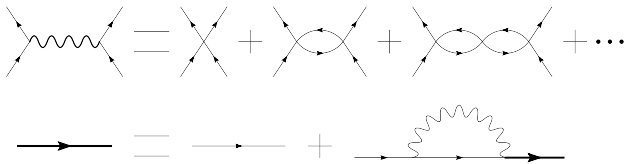, width=.46\textwidth}
\caption{The upper diagram defines the 
collective modes, i.e., `para-$\sigma (\pi)$' mode, which are
composed of the quark-antiquarks in RPA
and denoted by the wavy line. The lower diagram defines the quasi-quark
 which is dressed with the collective modes;
the thick and thin straight lines represent the dressed and free
quark, respectively. 
}
\label{selfe}
\end{center}
\end{figure}

The existence of  the collective modes composed of quark and anti-quark
in turn  modifies the quark properties.
Thus the problem becomes a self-consistent one where
the collective modes should be constructed with the 
dressed quarks and anti-quarks.
However, we take  in this exploring work 
the non-selfconsistent approach where the
collective modes are constructed by the undressed quarks and 
anti-quarks;
namely, the quark self-energy $\Sigma^R(\bm{p},\omega)$ is 
evaluated in the RPA as follows,
in the imaginary time formalism,
\be
  \tilde{\Sigma}(\bm{p},\omega_n) =
  -T\sum_{m}\int\frac{d^3 q}{(2\pi)^3} 
  \mathcal{D}(\bm{p}-\bm{q},\omega_n-\omega_{m}) 
  \mathcal{G}_0(\bm{q},\omega_{m}),
\ee
where
$\mathcal{D}(\bm{p},\nu_n)=\mathcal{D}_\sigma(\bm{p},\nu_n)
+3\mathcal{D}_\pi(\bm{p},\nu_n)$.
Fig.~\ref{selfe} is the diagrammatic expression 
for the quark Green function.
After the summation of the Matsubara frequency and the
analytic continuation, $i\omega_n\to\omega+i\eta$,
we obtain the quark self-energy in the real time,
\bea
  \lefteqn{\Sigma^R(\bm{p},\omega)}\nonumber \\ 
  &=& 2\int\frac{d^4q}{(2\pi)^4}
  \Lambda_+(\bm{q})\gamma^0
  \frac{{\rm Im}D^R(\bm{p}-\bm{q},q_0)}{q_0-\omega+|\bm{q}|-i\eta}
  \nonumber \\ && \times
  \left[ ( 1+n )( 1-f ) + nf \right]
  \nonumber \\
  & & +2\int\frac{d^4q}{(2\pi)^4}
  \Lambda_-(\bm{q})\gamma^0
  \frac{{\rm Im}D^R(\bm{p}-\bm{q},q_0)}
  {q_0-\omega-|\bm{q}|-i\eta}
  \nonumber \\ && \times
  \left[ ( 1+n )f + n( 1-f ) \right]
  \label{self-ene1}
\eea
with $D^R(\bm{p},\omega)={\cal D}(\bm{p},i\nu_n)|_{i\nu_n\to \omega+i\eta}$,
the projection operators 
$ \Lambda_\pm(\bm{q}) = (1\pm\gamma^0 \bm{\gamma}\cdot\bm{q}/|\bm{q}|)/2 $,
and the Bose and Fermi distribution functions
$ n = ( \exp(q_0/T) - 1 )^{-1} $ and 
$ f = ( \exp( |\bm{q}|/T) + 1 )^{-1}$.

Using the projection operators $\Lambda_\pm$,
the retarded quark Green function 
is expressed as
\be
  G^R(\bm{p},\omega)=
  \frac{\Lambda_+(\bm{p})\gamma^0}{\omega-|\bm{p}|-\Sigma^R_+ +i\eta}+
  \frac{\Lambda_-(\bm{p})\gamma^0}{\omega+|\bm{p}|-\Sigma^R_- +i\eta},
  \label{g-r1}
\ee
with $\Sigma^R_\pm(\bm{p},\omega) 
= (1/2) {\rm Tr}_D [ \Sigma^R \gamma^0\Lambda_{\pm}(\bm{p}) ]$
where Tr$_D$ denotes the trace over the Dirac index.
Each term in Eq.~(\ref{g-r1}) defines the 
quasi-quark and quasi-antiquark spectral functions, 
$
  \rho_\pm(\bm{p},\omega) 
  = -(1/\pi)
  \textrm{Im} [ 
  {\omega\mp|\bm{p}| - \Sigma_\pm^R(\bm{p},\omega)} ]^{-1},
$
respectively.
Remarks are in order here: A (anti-)particle number at finite
$T$ may be supplied either by a genuine (anti-)quark or by 
a `hole'  of the thermally-excited
  anti-quarks(quarks)\cite{Weldon:1989ys}, which 
implies that both the states or their mixed states
can  contribute to $\rho_+$($\rho_-$).

For later convenience,
we define the quasi-dispersion relations $ \omega = \omega_\pm( \bm{p} ) $ 
 as the zero of the real part of the inverse of the Green function;
$
\omega_\pm(\bm{p}) \mp |\bm{p}|
-{\rm Re} \Sigma^R_\pm(\bm{p},\omega_\pm(\bm{p})) = 0.
$
This quasi-dispersion relations will be found  useful as
an eye-guide of the peaks of $\rho_\pm$.
One should, however, be warned  that $\omega_\pm(\bm{p})$ 
does not necessarily represent the
real excitation spectrum when the imaginary part of the Green function is
large, since it is only 
a zero of the real part of the inverse of the Green function.
The physical dispersion relation should be identified
as the peak position of the energy $\omega$ as the function
of the momentum.

\section{Numerical results and discussions} \label{discuss}

\begin{figure*}
\begin{center}
\begin{tabular}{ccc}
$\varepsilon = 0.1$ & $\varepsilon=0.2$ & $\varepsilon=0.5$ \\
\epsfig{file=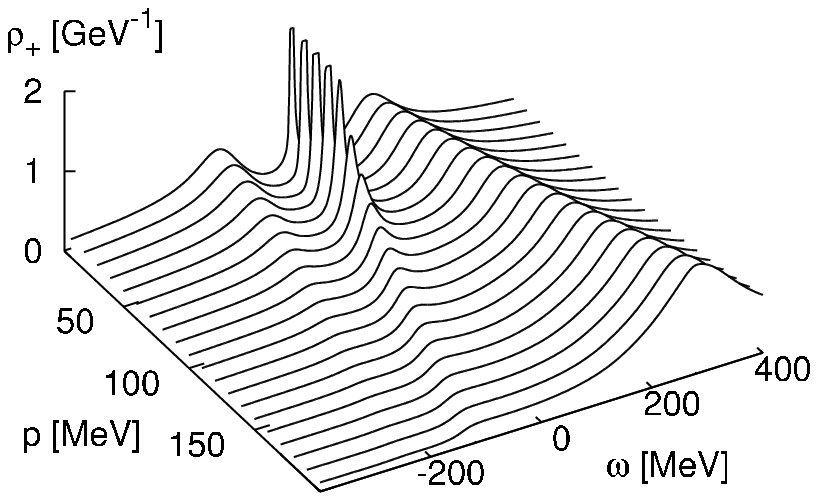, width=.32\textwidth} &
\epsfig{file=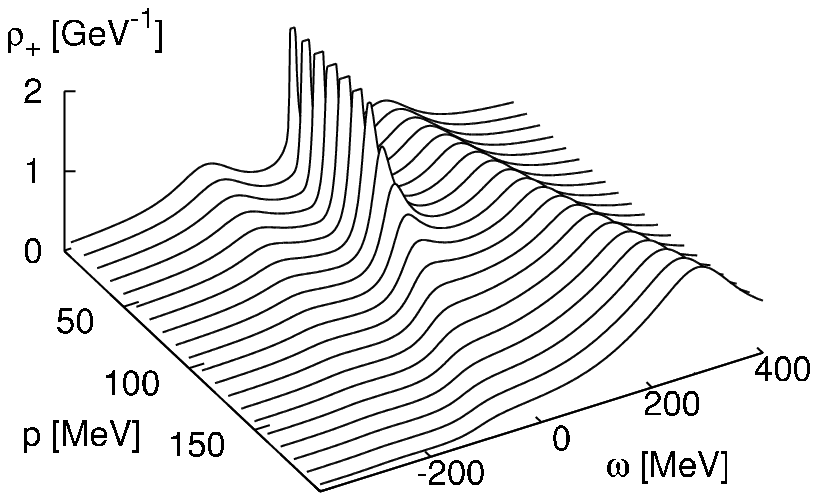, width=.32\textwidth} &
\epsfig{file=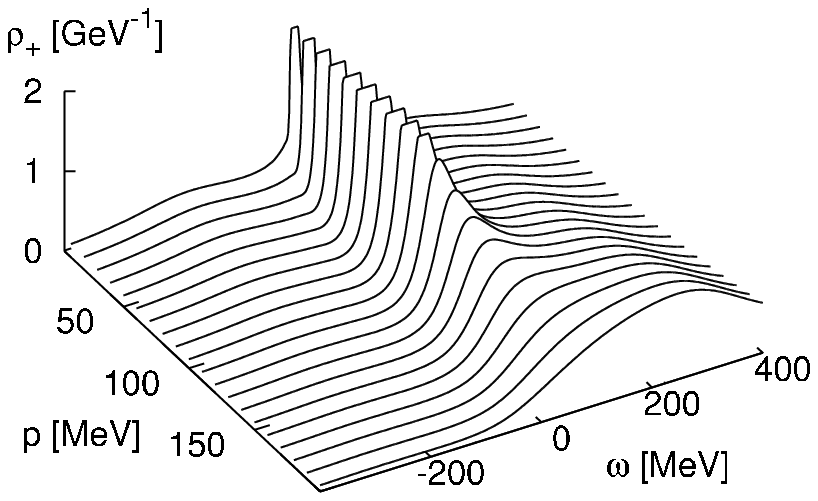, width=.32\textwidth} \\

\epsfig{file=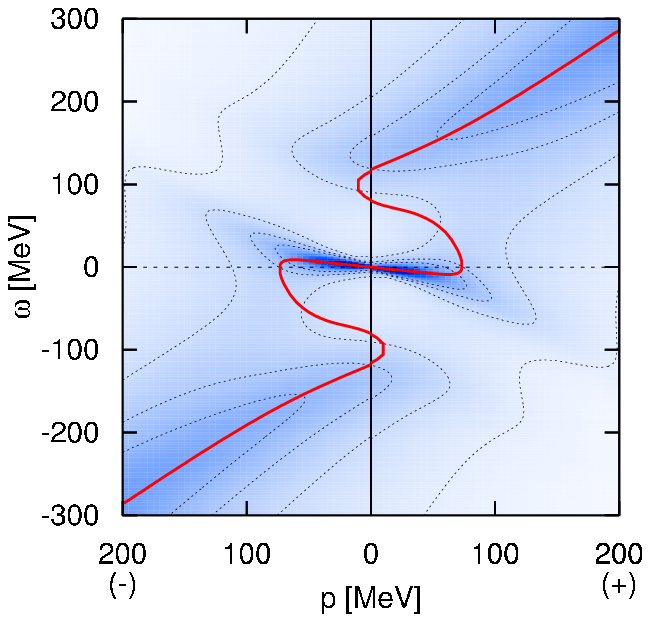, width=.32\textwidth}&
\epsfig{file=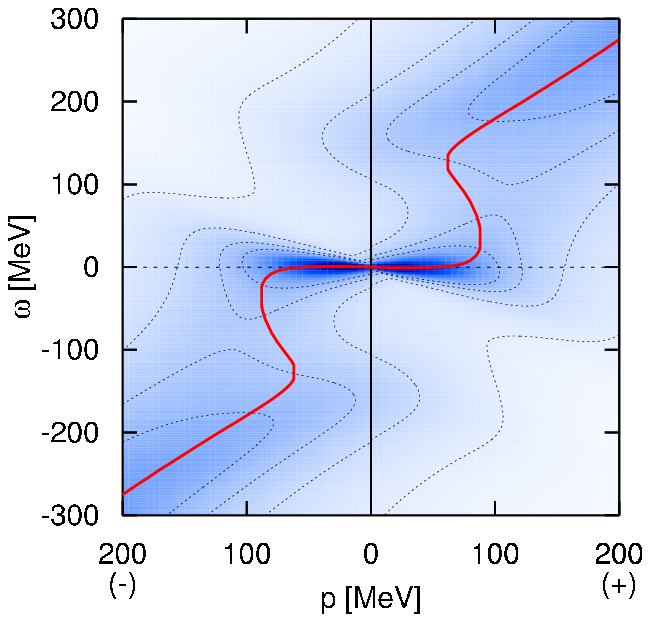, width=.32\textwidth} &
\epsfig{file=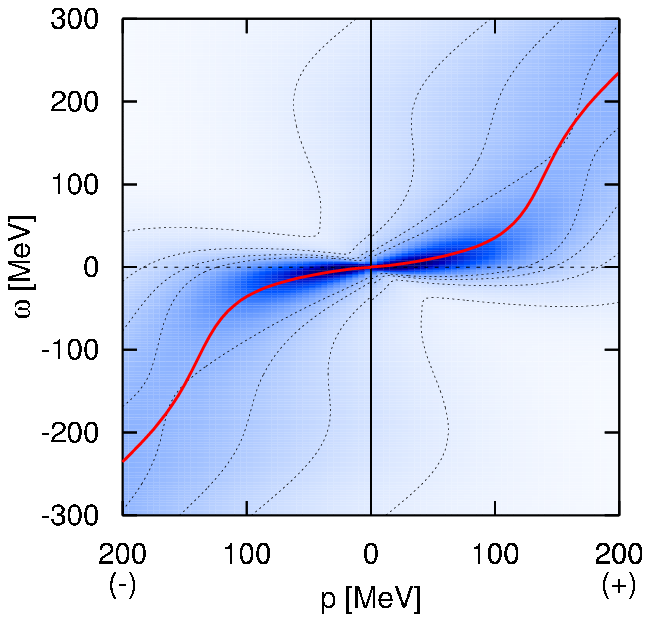, width=.32\textwidth}
\end{tabular}
\caption{
The upper panels
show the quark spectral function
$\rho_+(\bm{p},\omega)$ for the reduced temperatures
$\varepsilon = 0.1,0.2$ and $0.5$.
The lower panels  show
the dispersion relations $ \omega=\omega_\pm(\bm{p}) $ together
with the contour map of $\rho_\pm( \bm{p},\omega )$:
$\omega_+$ ($\omega_-$) and $\rho_+$ ($\rho_-$) are shown in
 the right (left) half of the figures; notice the direction of the
momentum scale in the left half plane is opposite to that of the right
half plane.
}
\label{fig:spct}
\end{center}
\end{figure*}

Now we shall show the numerical results of
the single-quark spectral function $\rho_\pm( \bm{p}, \omega )$
for several temperatures above $T_c$. 

In the upper panels of Fig.~\ref{fig:spct}, 
we show the quark spectral function
$\rho_+(\bm{p},\omega)$ for the reduced temperature
 $\veps  \equiv (T-T_c)/T_c = 0.1, 0.2$ and $0.5$.
We  remark that the anti-quark part $\rho_-$  of the spectral
function can be figured out  by the symmetric property,
$\rho_-(\bm{p},\omega)=\rho_+(\bm{p},-\omega)$.
The upper far-left panel shows the 
spectral function $\rho_+(\bm{p},\omega)$ for $\varepsilon=0.1$
 where the system is closest to the
critical point in the present three cases
and the soft modes are well developed as we have just seen
 in Fig.~\ref{fig:softmode}: 
One can see a clear three-peak structure in the spectral function. 
The detailed peak structure is seen in the contour map of the 
spectral function shown in the lower panel in the far-left.
One finds  that the quasi-dispersion relation 
$\omega_{\pm}(p)$   gives an approximate eye-guide of the peak position
of the spectral function; notice that
the dispersion relation for small momenta
$\bm{p}$ has a negative slope in the negative-energy region.
We remark that although 
$\omega_{\pm}(\bm{p})$ has an
acausal  `back-bending'-region in the $(p, \omega)$-plane,
 it does not correspond to 
any peak of the spectral function and hence
has  no physical significance. 

The  upper-middle panel of Fig.~\ref{fig:spct}
shows that the  three-peak structure still barely exists
even at higher temperature with  $\varepsilon=0.2$;
the would-be peak in the
negative energy region has turned to be a `bump' with a smaller strength.
The contour map of the
spectral function and the
quasi-dispersion relation in the lower panel show that 
the dispersion relation around the origin is somewhat shifted
upward and becomes almost a constant as a function of the momentum.
The far-right panels for $\veps=0.5$ show that when 
$T$ is raised further well above $T_c$, 
the three-peak structure disappears 
completely and the quark spectral function has a single peak
as the free quark system has, 
although the quark dispersion relation is modified at small momenta.

\begin{figure}[ht]
\begin{center}
\epsfig{file=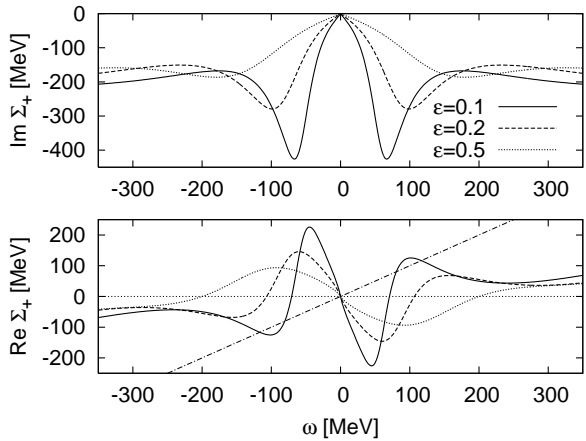, width=.46\textwidth}
\caption{The imaginary and real parts of the quark self-energy, 
Im$\Sigma_+^R(\bm{0},\omega)$ and Re$\Sigma_+^R(\bm{0},\omega)$.}
\label{fig:sigma}
\end{center}
\end{figure}

\begin{figure}[t]
\begin{center}
\epsfig{file=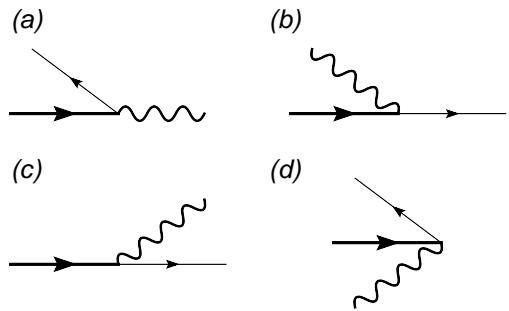, width=.37\textwidth}
\caption{Parts of the physical processes which the 
Im $\Sigma_+^R(\bm{p},\omega)$ describes.
The other parts are the inverse processes of the above.
The thick solid lines represent the quasi-quarks with $(\bm{p},\omega)$,
the thin solid lines the on-shell free quarks 
and the wavy lines the soft mode.
}
\label{fig:dcypr}
\end{center}
\end{figure}

We have now seen that the three-peak structure  
is the most characteristic feature in the quark spectra 
caused by the coupling with the soft modes.
To understand the mechanism of the appearance of the
three-peak structure in the quark spectral function,
we show the imaginary and real parts of $\Sigma_+^R( \bm{p}=0,\omega )$ 
for several values of $\varepsilon$
in the upper and lower panels in Fig.~\ref{fig:sigma}.
From the upper panel, one sees that
there develop two peaks in $|{\rm Im}\Sigma_+^R|$ 
in the positive- and negative-energy regions
at small temperatures,
with the peak positions moving 
toward the origin as $T$ is lowered to $T_c$.
It is found that 
the peaks in Im$\Sigma^R_\pm( \bm{p},\omega )$ in the positive and
negative energy regions essentially correspond to
the decay processes shown
in Fig.~\ref{fig:dcypr}(a) and (b), respectively,
 where the wavy lines represent the soft modes.
As noticed before, 
when  $T$ is close to $T_c$,
the soft modes acquire the character of the well-defined 
elementary excitation with a mass $m^*_\sigma(T) $ and a small width,
 and
thereby the system
can be described by a Yukawa theory with a
massless quark and a boson with a finite but small mass. 
In fact, we have confirmed that the essential
features obtained here, including the three-peak structure in 
the quark spectral function can be reproduced by such a Yukawa model
\cite{ykw}.
The processes in Fig.~\ref{fig:dcypr}(a) and (b)
are also interpreted as a Landau damping  of a quark state
due to the collision with the thermally excited soft modes with
the dispersion relation $\omega=\omega_s(\bm{p};T)$.
One should notice here that the incident anti-quark line 
in Fig.~\ref{fig:dcypr}(a) may
describe a thermally excited antiquark, which disappears after the 
collision with the soft modes. 
But the disappearance of the anti-quark means the
creation of a hole in the anti-quark distribution\cite{Weldon:1989ys}.
The point is that an `anti-quark hole' has a positive
quark number.
Fig.5(b) describes the decay process of a quasi-quark
state which is a mixed state of quarks and antiquark-holes 
to an on-shell quark via a collision with the soft modes.
These processes induce a quark-`antiquark hole' mixing, which constitute
the physical states in the system with a modified spectrum.
Notice that  the closer the $T$ to $T_c$, 
more significant the Landau damping processes shown in
Fig~\ref{fig:dcypr}, because the thermally excited soft modes
are expected to be abundant.

The mixing mechanism of quarks can be described 
in terms of the notion of {\em resonant
scattering} as in the case of the (color-)superconductivity
\cite{JML97,Kitazawa:2005pp}, although a crucial difference
arises owing to the different nature of the soft modes.
In the case of the superconductivity,
the precursory soft mode 
is diffusion-mode like and has a strength around $\omega=0$.
A particle (electron or quark) is scattered by the soft mode and 
creates a hole, and vice versa, whose process is called 
a resonant scattering.
The resonant scattering with the soft mode induces the mixing between
a particle and a hole, and  thereby giving rise to a gap-like
structure in the fermion spectrum around the Fermi energy; 
correspondingly, the imaginary part $|{\rm Im}\Sigma^R|$ of the 
quark self-energy has a single peak around 
the Fermi energy $\omega=0$\cite{Kitazawa:2004cs2,Kitazawa:2005pp}.
In the  present case,
the soft modes are propagating modes and
 have a strength at finite $\omega=\pm \omega_s(\bm{p};T)$ 
both in the positive- and negative-energy regions, and 
 hence the resonant scatterings of the quarks with the
chiral soft modes give rise to two peaks in
$|{\rm Im}\Sigma^R|$ at finite energies roughly
 of the order of $m^*_{\sigma}(T)$
and  induce a mixing   between a quark and an
antiquark-hole and  between an antiquark and a quark-hole, respectively.
Thus the two gap-like structures in the quark spectrum are formed
at the positive- and negative-energies.

The modified quark spectra can be graphically obtained from
 the lower panel of Fig.~\ref{fig:sigma} where
 Re$\Sigma^R_+( \bm{0},\omega )$ is shown:
One can see there exist two regions of $\omega$ where
Re$\Sigma^R_+( \bm{0},\omega )$ shows a steep rise.
These two regions correspond to that in which 
$|{\rm Im}\Sigma^R_+|$ has a peak
and the steepness becomes 
more significant as $T$ is lowered toward $T_c$.
This behavior of Re$\Sigma^R_+$ is naturally understood 
from the behavior of Im$\Sigma^R_+$
using the Kramers-Kronig relation.
Because the quasi-quark dispersion  $\omega_\pm(\bm{p})$
at the  vanishing momentum
is  the solution of $\omega - \Sigma_\pm^R (\bm{0},\omega) = 0$,
they are given graphically by the crossing 
points of  Re$\Sigma_\pm^R(\bm{0},\omega)$ and $\omega$ 
denoted by the straight 
dash-dotted line in the lower panel of Fig.~\ref{fig:sigma}.
One sees that there eventually appear five crossing points
for $\varepsilon=0.1$ corresponding to the five solutions
of $\omega_+(\bm{p}=\bm{0})$ in Fig.~\ref{fig:spct}.
The second and forth solutions of them, however,
correspond to the peak of $|{\rm Im}\Sigma_+^R(\bm{p},\omega)|$ and hence
there  appear no peaks in the spectral function in this region.

As $T$ approaches $T_c$ closer than those in Fig.~\ref{fig:spct},
$m^*_\sigma(T)$ and hence  the ratio $m^{*}_{\sigma}(T)/T$
 become vanishingly small,
the system may become identical in effect 
with an extremely high-$T$ system where
the hard-thermal loop(HTL) approximation is valid.
In fact, in such a critical region,
 the two peaks of Im$\Sigma^R_+$ 
in the positive- and negative-energy regions
come closer and merge together effectively 
with the peak at $\omega=0$ and thereby
only a single resonant scattering around $\omega=0$ is seen
in the limit of $T\rightarrow T_c+$. This means that there
will be eventually only two peaks
originated by the scattering of a massless quark with
a massless boson at finite $T$. 
Thus the resultant quark spectrum 
at $T\rightarrow T_c+$ continuously approaches
that 
obtained in the HTL approximation
in the Yukawa theory\cite{Baym:1992eu}.
Although not explicitly shown in the present paper because of
the lack in space,
our numerical calculation
 shows that this is indeed the case:
The strength of the peak around $\omega=0$ becomes weaker 
as $T$ is lowered toward $T_c$,
while the width of the two peaks which have a  thermal mass 
becomes sharper.
The latter two peaks eventually become the 
quasi-particle peak and the plasmino peak when $m^*_{\sigma}(T)/T \to 0$, or 
equivalently, $T \to T_c+$.

\section{Brief summary and Concluding remarks} \label{conclusion}

We have investigated the quark spectrum at $T$
near but above the critical temperature of the chiral phase transitions
taking the effect of the fluctuations of the chiral
order parameter (para-$\sigma$ and para-$\pi$ modes) into account.
We have shown that 
for $\varepsilon\equiv (T-T_c)/T_c \lesssim 0.2$
the quark spectrum has a three-peak structure 
at low frequency and low momentum, for which
 the growth of the two peaks in Im$\Sigma^R_\pm$
for lower $\omega$ near $T_c$ is crucial.
We have elucidated that the mechanism for realizing the two
peaks  can be
understood as a formation of gaps owing to
the  mixing between a quark(anti-quark) and 
a hole of the thermally-excited antiquarks(quarks); the mixing is induced 
by the {\em resonant scattering} of the quarks off the propagating
soft modes with small but finite energies near $T_c$.

The present discussions to account for the mechanism of the formation of 
the three-peak structure have taken it for 
granted that  
the soft modes which are composite system
of quark-antiquarks acquire the character 
of the well-defined elementary excitation
when  $T$ is close to $T_c$.
It would be helpful for elucidating
 the essential mechanism quantitatively 
to consider a Yukawa model at finite $T$, in which 
the mass of the boson field is varied by hand\cite{ykw},
where we have confirmed that the three-peak structure of the
quark spectral function emerges in such a Yukawa model.
Moreover, our preliminary calculation\cite{ykw} shows that 
the three-peak structure 
is obtained even if the bosonic modes are  vector or axial-vector fields,
instead of the scalar (pseudoscalar) field.
This result suggests
that any bosonic mode in the QGP phase,
whose mass is comparable with $T$, can give rise to
the three-peak structure of the quark spectral function:
Notice that the peaks of the spectral function correspond to 
complex poles of the quark Green function, which are gauge
invariant!
The detailed analysis of the quark spectrum 
will be given elsewhere\cite{ykw}.

We have adopted the chiral limit in this exploring work for simplicity.
It is instructive to investigate the quark spectrum 
for finite quark mass.
In this case, the chiral transition becomes crossover
and the mass of the $\sigma$ mode does not vanish in any $T$.
We expect, however, that the three-peak structure of the quark 
will be seen even in this case 
because the three-peak structure is seen in rather wide range of $T$.
It is also instructive to explore the quark spectrum
at finite chemical potential,
in particular near the tri-critical point and/or 
critical end-point of the chiral transition.
In this work, we have also 
employed the non-selfconsistent approximation.
It is of course more desirable 
to adopt more sophisticated approximation incorporating the
self-consistency between the soft modes and the quasi-quarks,
especially near $T_c$.

Some authors \cite{Babaev:2000fj,Castorina:2005tm} suggest
the importance of the phase fluctuation
of the chiral order parameter  on the properties of the QGP above $T_c$.
It would be interesting to explore possible effects
of  the phase fluctuation on the quark spectrum.

{\bf \hspace*{-1.3em} Acknowledgments}
\vskip 0.1em
We thank H. Fujii for useful discussions.
Y. N. thanks M. Harada and K. Yamawaki for useful comments.
Y. N. was supported by the 21st Century COE Program of
Nagoya University.
M. K. is grateful to Dirk H. Rischke and other members of Frankfurt
university for the warm hospitality extended to him during his stay there.
He is supported by Japan Society for the Promotion of Science for
Young Scientists.
T.K. is supported by Grant-in-Aid
for Scientific Research by Monbu-Kagakusyo
(No. 17540250).
This work is supported by the Grant-in-Aid for the 21st Century COE 
``Center for Diversity and Universality in Physics" of Kyoto
University.

\end{document}